**CHAPTER 26**

# Software Engineering Through Community-Engaged Learning and an Inclusive Network

**Nowshin Nawar Arony, Kezia Devathasan, Ze Shi Li, and Daniela Damian**

Retaining diverse, underrepresented students in computer science and software engineering programs is a significant concern for universities. In this chapter, we describe the INSPIRE: STEM for Social Impact[1] program at the University of Victoria, Canada, which leverages the three principles of self-determination theory – competence, relatedness, and autonomy – in the design of strategies to empower women and other underrepresented groups in using software and other engineering solutions to approach sustainability, community-driven problems. We also describe lessons learned from a first successful year that involved over 30 students, 6 community partners (sustainability problem owners), and over 20 industry and academic mentors and reached out to more than 200 solution end users in our communities. Finally, we provide recommendations for universities and organizations who may want to adopt our approach.

---

[1] https://inspireuvic.org/

In the program 24 diverse students (in terms of gender, sexual orientation, ethnicity, academic standing, and background) divided into six teams paired with six community partners worked on solving society-impactful problems and developed solutions for a number of respective community partners. Each team was supported by an experienced upper-year student and mentors from industry and community throughout the program. The experiential learning approach of the program allowed the students to learn a variety of soft and technical skills while developing a solution that has a social and/or environmental impact. Having a diverse team and creating a solution for real end users motivated the students to actively collaborate with their peers, community partners, and mentors resulting in the development of an inclusive network. A network of like-minded people is crucial in empowering underrepresented individuals and inspiring them to remain in the computer science and software engineering fields.

## Introduction

Computer science and software engineering university programs have long suffered from a lack of diversity where recruitment and retention of students from underrepresented groups are challenging [3]. Working on society-impactful projects has the potential to motivate women and other underrepresented individuals to continue in Science, Technology, Engineering, and Mathematics (STEM) as they are often drawn toward care-oriented and humanistic careers [4, 6]. Experiential learning, a method employing learning through working [8], has proven to increase confidence and motivation in students to continue and persist to graduation [9].

In recent years, some universities have begun launching initiatives and protocols regarding increasing equity, diversity, and inclusion (EDI) in engineering programs to train faculty, staff, and even teaching assistants [16]. These protocols are a means to help reduce the potential harm that could inflict on a student. In Part 4, Chapter 15, "Beyond Classroom: Making a Difference in Diversity in Tech," the Czechitas initiative has been described that supports and trains girls in computing education to succeed in tech careers. The community developed through this initiative resulted in the increase of women participants in their following year. Having a supportive network to grow develops a sense of belonging in underrepresented individuals. Furthermore, in Part 5, Chapter 25, "Effective Interventions to Promote Diversity in CS Classroom," the authors describe how "interest" enhances the learning process and engraves motivation in girls. Engaging students in projects with social and/or environmental causes and allowing them to interact with real clients can increase interest and confidence of students. Hence, we launched INSPIRE that aims to foster EDI through community-engaged experiential learning for underrepresented students and support them through a network of like-minded individuals. We incorporate design thinking methodology [10] to facilitate students' learning experiences as they collaborate with local nonprofit and for-profit organizations, which we refer to as "community partners."

In this chapter we share the student experiences working in this program. The preliminary goals of the program included *(1) developing an inclusive network of individuals, (2) motivating students through empowerment, (3) providing experience with realistic and impactful problems through community engagement, and (4) learning to work in diverse teams*. Through analyzing program participants' experiences, we identified important lessons and present recommendations that may be beneficial for future practitioners and educators implementing diversity-centered experiential learning opportunities in other institutions.

## Program Overview

The INSPIRE program offered a four-month-long summer internship position to science and engineering students at the university to work in teams to solve society-impactful problems for real clients. The program was open to undergraduate and graduate students from varying science and engineering degrees ranging from first year to final year. The students were required to work 40 hours/week and received co-operative education (co-op) credits for their work in the program.

The fundamental idea of the program originated from the principles of self-determination theory [14], which describes the factors that contribute to different types of motivation rooted in three basic needs: *competence*, a need to be able to effectively handle the environment; *relatedness*, a need to have close bonds with others; and *autonomy*, the freedom to make one's own choices. Intensive training was provided for all aspects of the students' projects to provide learners a sense of competence. An emphasis was placed on team bonding activities (e.g., ice-breaking discussions, giving constructive feedback to each other, playing games in groups) and group activities (e.g., hiking, beach walk) with all teams together to create a sense of relatedness in the program network. Finally, while students received extensive training, the details of their projects were not micromanaged by the program management team, providing a sense of autonomy within an otherwise structured program.

The program secured sponsors from across Canada that included IBM Canada Advanced Studies, Riipen, Redbrick, Checkfront, McElhanney, PBX Engineering, WSP, iWIST (Island Women in Science and Technology), Viatec, Actua, Inter-cultural Association of Greater Victoria, KWL Consulting Engineers, Axolotl Biosciences, Animikii, and Ocean Networks Canada. Many of the industry partners further offered mentorship to the students; as such, we had over 20 industry mentors supporting the students throughout the four months in various project-related aspects. Some of the mentors were located in countries other than Canada like China, Pakistan, Brazil, and the United States.

Along with that, we had six local community partners: Swan Lake Christmas Hill Nature Sanctuary, Greater Victoria Coalition to End Homelessness, Victoria Brain Injury Society, NatuR&D, local schools in the Greater Victoria area (Claremont Secondary, GNS, and Ecole Victor-Brodeur), and Redbrick. The community partners were local to Victoria as the students needed to work with the clients in person. Many of the community partners were nonprofit organizations who actively engaged in helping those in the community that are most vulnerable.

Finally, the program timeline consisted of eight phases encompassing four months. Table 26-1 shows the different phases during the four months and the tasks involved.

**Recruitment and project selection:** To maximize the reach of potential participants, an open call was sent through the university platform (i.e., email, social media, etc.) four months prior to the launch of the program. Since the program was based on experiential learning and empowering students, the recruitment did not include any previous course experiences for the student teams. However, a separate call for upper-year experienced students, preferably fourth- or fifth-year undergraduate or graduate students, was made in parallel who would be supporting the students with administrating logistical or teamwork issues. Over 50 students responded who were interviewed through a two-step filtering process that was designed to test different soft skills such as primary communication skills, conflict resolution, and leadership.

The first step consisted of a team activity to test each student's ability to (1) work in a team, (2) overcome any conflict that would occur in that short time span, and (3) self-organize the team decisions. Students were broken into teams of five to six students and tasked with completing a project that addressed a hypothetical problem and constraints, which would require students to demonstrate their skills in these three areas. The second phase of the recruitment was an interview process where each student was interviewed individually on different situational questions, for example, how would they behave under a conflicting situation, how would they deal with failure, or what would their approach be when conversing with the community partner. At the end of this recruitment process, 30 students were recruited in the program.

*Table 26-1.  Timeline of the program*

| Month | Phases | Tasks Involved |
|---|---|---|
| May | Problem Definition | Students were provided with interpersonal skills training like equity, diversity, and inclusion (EDI), professional conduct and communication with clients, and conflict resolution, project-specific training, design thinking, as well as technical skills training like web frameworks, version control, and Agile. Students were required to finish defining the initial problem. |
| | Problem Planning and Framing | Students met with their respective community partners and end users to further plan with them and start framing the problem based on collected user data (interviews, focus groups, surveys). |
| June | Problem Validation and Early Prototyping | Students revisited the clients to validate their findings in order to start prototyping. They further started developing an initial prototype. |
| | Midterm Presentation | Students showcased their work to the public and received feedback from professionals regarding their project. |
| July | Validation of Prototype | The students reiterated their prototype with the clients before finalizing it for implementation. |
| | Solution Development | Students started the solution development and were required to learn the necessary technology. |
| August | Continued Solution Development | Fully dedicated to the development of the solution, students worked in collaboration with the clients to make sure the clients were satisfied with the product. |
| | Finalizing Solution Development and Documentation | Students began wrapping up the project and creating proper documentation for the clients so that the clients could take this forward and utilize it in their community. |

Similarly, an open call for the community to propose projects was made through the community-engaged learning (CEL) department at the university. We selected a total of six projects as they were the most pertinent for the students and program. Not only did each project address a pressing social or environmental challenge affecting the broader community but each community partner also committed to mentoring the students in project-specific training such as dealing with brain injury patients and norms of engaging with vulnerable clients.

**Team formation:** Out of the 30 students, 24 students were placed in a group of six teams, and each team was assigned a senior experienced student to support them. Prior research showed that working in a project that resonates with one's value gives a sense of motivation to work on the project [13]. Thus, each student was asked to provide their top three project preferences as well as their academic skills, experiences, and project requirements, and then each student was placed in a team accordingly. Each team ended up with diverse members in terms of gender, sexual orientation, ethnicity, academic standing, and background. Moreover, each project had a diverse set of end users, making the projects more challenging and motivating.

**Demographic diversity:** Among the 24 students, 10 identified as female, 12 identified as male, and 2 preferred not to disclose their gender. Furthermore, 19 students came from the undergraduate level and 5 from the graduate level. The students were further diverse in terms of (1) academic background as they came from computer science, software engineering, electrical engineering, mechanical engineering, biomedical engineering, physics, chemistry, and business and (2) ethnicity, including South Asian, East Asian, Black, Arab, Hispanic, Indigenous, and White. Out of the six experienced students, there were five female and one male student, and each of them had different engineering and science backgrounds. The teams were further matched with industry mentors who guided them on different social and technical issues. Therefore, while each project team had a diverse blend of experiences, skills, and perspectives, all team members had an equal opportunity to work on a project that they were deeply motivated to work on.

# Projects

The four-month program was both an accomplishing and turbulent journey for all the students, as the teams had to overcome various adversities. The first month was heavily dedicated to training and learning about the project-specific requirements. Since all the community problems appeared to require technological solutions, the students had to learn different technical skills including programming languages, frameworks, version control, databases, API integration, PCB design, geographic information systems, and many more depending on their project.

*Table 26-2.* *Summary of the projects and the solutions*

| Project Name | Community Problem | Solution |
| --- | --- | --- |
| Swan Lake Christmas Hill Nature Sanctuary | Due to the increase in the number of people visiting a local nature sanctuary, preserving and maintaining specific areas is becoming difficult. | An IOT monitoring system to track and visualize where visitors are trekking in the park |
| Greater Victoria Coalition to End Homelessness | Women+ fleeing violence and facing homelessness encounter difficulty finding safe and appropriate support, services, and housing. | A website that allows support workers to easily find up-to-date available emergency housing and services for women+ fleeing violence |
| Victoria Brain Injury Society | Nonprofits supporting brain injury survivors lack a centralized, accessible hub to provide patients with relevant services. | A mobile application with a custom interface directing brain injury survivors toward necessary and appropriate services and support |
| Carbon footprint awareness for teens | Youth lack the motivation to take climate action due to inadequate knowledge and inspiration. | A gamified classroom app that is designed to help teenagers learn and take action about their personal carbon emissions |
| The Resilient Urban Systems & Habitat Initiative | Existing climate change data is disorganized and fails to provide informed guidance on potential climate health. | An interactive website that centralizes and reports information about regional climate change vulnerabilities |
| Carbon impact of web browsing | Digital activities are part of everyday life, but people are unaware of the carbon impacts of browsing the Internet. | A web application that accurately calculates the carbon impact of web browsing |

The teams were introduced to IBM's design thinking [10] and Agile [1]; as such, they extensively utilized these processes in their project management, software development, and requirement elicitation. Due to the heavy emphasis on experiential learning, the students mostly developed skills through implementing these skills during the software development process. In addition, the students learned a plethora of soft skills, some of which were unique to their project due to having specific clientele. For example, a team of students working with patients who had acquired brain injury required training on methods of interaction with such patients. Gathering requirements from vulnerable groups is sometimes difficult; hence, learning such skills was a significant part of their project. All six projects had pressing community problems, which allowed the students to explore the phases of design thinking [10], learn, and develop solutions that could cater toward the respective end users. Table 26-2 summarizes a brief description of the community problems and the final solutions developed by the teams.

All six projects had unique experiences in terms of overcoming different kinds of obstacles and pivots. However, by the end of the four months, they were left with the sense of achievement and learned different aspects of working in a software development team. Comparing the program with a course, one of the students said: *"This is so much better than taking a course. I think because in a course, the projects feels contrived. And I don't feel like the end result actually does anything. I mean, you learn through it. But it's, that's the intention is learning. Where here learning is not the only intention. It's about, you know, building community and making, you know, building interpersonal skills and really setting ourselves up for the future while also making a product that actually will go out into the world and do some good."* This quote is a perfect summary of the program's goal of creating a network of like-minded people and contributing to their success through motivation, empowerment, mentorship, and curating a safe space in the community.

## Research Methods

Over a period of four months, we collected reflections and conducted focus group interviews with students and community partners to improve our program for the next iteration. Workshops were provided to help students reflect on their learning outcomes from working on real-world problems in diverse teams. As part of the deliverables, students were required to write weekly individual and team reflections based on their project development and teamwork experience. Two focus group interviews were conducted with each team, and community partners were also interviewed to gather their perspective. We collected approximately 300 individual reflections, 90 team reflections, 12 focus groups, and 5 community partner interviews. The collected data was analyzed through Braun and Clarke's [5] six-step thematic analysis method that includes (1) familiarizing with the data, (2) generating initial codes, (3) searching for themes, (4) reviewing potential themes, (5) defining and naming themes, and (6) producing the report. An external member outside the research team reviewed the findings to avoid assumptions and biases. The research team used an iterative discussion cycle and peer debriefing process to extract key themes from the collected data, which we present in the form of the lessons learned in the following section.

## Lessons Learned and Recommendation

This section highlights the lessons learned from the four-month-long projects. Working on a community project with *real clients* and *community problems* poses both rewarding and challenging experiences that students would not otherwise be exposed to until post-graduation employment. The students in our program were provided with a unique combination of working with real clients tackling real problems in diverse teams. During this time, we analyzed collected personal and team reflections as well as conducted focus groups with the students to understand their experience. We describe in the following the lessons based on this analysis, which after much reflection, we propose, can be supported pedagogically.

**Lesson 1 –** *Training on soft skills* should be emphasized in *building a successful network of diverse individuals in software engineering*: In a successful network, people are able to socialize and support each other outside of work and develop meaningful relationships. Previous research suggests that such nonprofessional relationships, even in organizational networks, are critical to companies employing research and development projects [14]. Soft skills are immensely important in building such a community, yet university students often lack the soft skills that are required to operate in the real workplace [2]. Thus, training in such area is critical, as a lack of adequate training can cause conflicts, customer dissatisfaction, and team fallout [11]. This program was the first exposure for many students to work with real clients; therefore, we provided them with a number of training sessions covering soft skills such as communication with clients; team management; professional conduct; equity, diversity, and inclusion (EDI); and leadership. This was done to ensure successful interactions between students and clients such that positive relationships between them could be built, thus facilitating the growth and strengthening of their networks.

Due to the unique end users each project catered toward, eliciting requirements from clients and end users while also behaving in a professional and respectful manner was likewise an important skill to learn. Positive interactions with potential end users also contributed to students learning to successfully expand and integrate into their own networks. As a result of the provided soft skills training, community partners were immensely satisfied with the students as one of them stated: *"I thought they were very organized and professional."* This suggests a willingness to interact with our students again, which confirms that a strong rapport has been established.

As such, the community partners further connected them to individuals who could help them with their project. Describing their experience building connections, one student said: *"The connections are amazing; we couldn't have had those connections and get in touch with them as quickly as possible without [our community partner]."* This quote implicitly indicates the importance of communication skills for building connections. To help students in communication development, we trained them with techniques such as writing down constructive feedback for team members. *"This exercise has made it clear to me that communicating constructive feedback is something that I need to learn, and I hope that I may learn how to do it kindly."* This confirms that open communication, while difficult to establish, is critical for effective relationship-building, thus justifying a need for soft skills training to help students build their own professional networks.

*Recommendations:*

- Dedicate and emphasize a significant time toward soft skills training to help students build connections and develop a supporting network.

**Lesson 2 –** *The right amount of guidance empowers students to balance autonomy and motivates them***:** While we encouraged our teams to be largely self-organized, this proved to be a delicate balancing act. We realized throughout the duration of this program that not enough autonomy would lead to feelings of being micromanaged. On the other hand, too much autonomy meant that students sometimes felt lost and were facing overwhelming uncertainty. We observed this closely in our cohort, with some students expressing a lot of anxiety and stress early in the semester when faced with independence: *"I would say, the very beginning with the whole trying to figure out research and stuff on the different keywords, I think we were all kind of in the* *same boat at that point, trying to like figure out what we're doing."* Work by Noll et al. [13] supports these feelings, speculating that individuals with lower competence, such as our students at the start of the term before learning new skills, will not benefit from high levels of autonomy. For this reason, we front-loaded substantial technical, soft skills, and EDI training in the program so that every student would receive some initial guidance in a variety of soft and technical skills to increase feelings of competence.

As the term progressed, the contrary was observed through the increase in competence and relatedness of students in the program. Throughout the term, students developed a variety of skills and had the opportunity to bond with their team members. Team members often helped each other overcome different challenges or navigate knowledge gaps leveraging their diverse backgrounds and skillsets. As the term progressed further and students further developed their competence, they expressed an appreciation for autonomy; one student says, *"We're given the space to come together and kind of figure it out what's needed, in my opinion, as a team to kind of figure out how to grow together."*

With the teams' progressing through changes in their own competence and relatedness, the teaching team required to adjust how much intervention was needed with the team's processes. We realized that students built up autonomy over and relied on our guidance less and less as the semester progressed. Thus, incorporate motivating the students to work in their own pace and empower them to succeed.

*Recommendations:*

- Provide adequate support to ensure students are not overwhelmed by autonomy and monitor student feedback to adapt if necessary.

- Consider changing skillsets of the students and adjust the level of support accordingly.

**Lesson 3 –** ***Proper structure enhances community-engaged collaboration between the students and the community partners*:** In this program, community partners' collaboration was crucial to the success of the experiential learning process. Students expressed feeling highly motivated as a result of working in a real project with the community partners. The students felt more accountable in producing a successful final product as this product would be deployed to the community. When comparing the program to a typical course, one student described, *"You just have an imaginary community partner or user requirements that doesn't change over time. It's just solid; they give you a problem statement, and you solve it."* However, they expressed that in this

459

program *"the element of getting real people [had] a big role, because what [they] build might actually end up saving lives."*

Despite real clients being so important to the students' motivation and success, the community partner's vision for the solution collided with the students' skillsets. For some of the projects, the scope was so large that the students were required to conduct extensive user research in the first two months to define the scope to a workable state as well as consider their own skills and knowledge to set the scope. Project 5, for example, had a somewhat unclear and broader scope as mentioned by one of the students: *"It's clear that the project's scope is significant and, at times, very daunting."* Thus, they negotiated the scope of their project with the community partner. The community partner for this project mentioned having to bring students back on track as oftentimes they would deviate. *"We seem to have been pivoting and spinning our wheels a little bit more … I do know that because of the scope of the existing problem. It was really huge. And to try and keep the team focused on just chunking out something small, as part of it, was a task in itself."*

As a solution for such expectation conflicts, the community partners expressed the need for more guidance (written guidelines) regarding how much their involvement should be. Students conveyed similar needs. To facilitate this improvement, one of the community partners suggested *"more touch points across groups, offering everybody a chance to get together or something."* They also expressed that it would have been helpful to *"hear from the other community partners to see how it's going for them."* Hence, more support is desired by both students and clients to make this collaboration more successful to build a supportive network.

*Recommendations:*

- Provide a guideline to the community partners to mitigate uncertainty regarding how they are expected to interact with students.

- Add more instruction and training in communication, negotiation, and scoping skills, since many students are engaging with community partners for the first time.

**Lesson 4 – *Through mentorship and EDI training, students learn to overcome challenges of working in a diverse team.*** Diversity has the potential to both benefit and hinder team performance [7]. One student described a frustrating experience in which they were encountering too many perspectives, saying, *"The most eye-opening* *thing to me is like, how we can have like, the same objectives but like, the same goals but like different ways and like solving the same problem."* However, the EDI training helped them realize the importance of different views in a team. The EDI training consisted of targeted workshops where a EDI trainer from the university explained the various facets related to equity, diversity, and inclusion.[2] The training consisted of several exercises that enforced critical thinking on understanding one's own privileges and the use of correct wordings while engaging with different people. The concept of empathy and equity was a key discussion in this training.

In the early stages, the students would often prioritize their own work as "important" over others, which would result in frustration for the other members. This highlights the teams going through the storming phase of Tuckman's [16] model of group development. However, with adequate mentoring from their industry mentors and instructional team, they soon realized that it was necessary to learn to discard personal biases for the betterment of the project. One student said, *"I have a process or method that I have developed on my own and naturally think it is the best and most efficient system ever, but it's clashing with these other three or four [team members]. I have never worked in a group setting before where our views and opinions could differ so significantly on such a small detail, mostly fascinating than something to be concerned about. I do see what they mean and try to understand why that is truly the best solution, most of the time it is, which is really cool to see how the collaboration worked to create the most efficient solution."*

As a result of the continuous guidance on practicing EDI, soon the students started leveraging their team diversity through efficient work distribution. Diversity was a prominent part of the program, and practicing inclusion in teams was constantly encouraged though training and mentorship. Hence, by the end of the program, the students realized how impactful working in a diverse team can be. *"I have found working on a diverse team enjoyable. It has given me the opportunity to learn new things, as everyone is in a different discipline and has different specialties. My ability to learn from my teammates is made possible through the team culture, which encourages asking questions, getting feedback from each member, and offering as much assistance as possible."*

---

[2] www.uvic.ca/equity/education/workshops/index.php

461

*Recommendations:*

- Include explicit EDI training throughout the course of the project and emphasize the benefits of working with diverse teammates.

- Be prepared to provide mentorship to students as conflict is expected from time to time.

## Conclusion

We presented a pioneering program that aimed to motivate students from underrepresented backgrounds to stay and succeed in computer science and software engineering through community-engaged experiential learning. Over the course of four months, the students first received soft skills and technical training from the instructor team. Consecutively, they engaged with the community partners in identifying and scoping the problem before conducting weeks of prototyping and solution validation. Solving the diversity problem is not a small feat, and we hope that our program design, lessons learned, and recommendations are useful for other universities and organizations looking to help tackle the issue in their own communities. The experiences described in this chapter represent the first step of our INSPIRE program aiming to help address diversity in computer science and software engineering. We summarize our lessons learned and recommendations as a number of takeaways (see Table 26-3) from our experience running the first cohort of INSPIRE.

*Table 26-3.* *Takeaways*

| Lessons Learned | Recommendation |
| --- | --- |
| Training on soft skills facilitates building a successful network of diverse individuals in software engineering. | – Dedicate and emphasize a significant time towards soft skills training to help students build connections and develop a network. |
| The right amount of guidance empowers students to balance the autonomy and motivates them. | – Provide adequate support to ensure students are not overwhelmed by autonomy.<br>– Consider how skilled the students are becoming and adjust the level of support accordingly. |
| Proper structure enhances the community engagement experience between the students and community partners. | – Provide a guideline to the community partners to mitigate uncertainty regarding their expected interaction with students.<br>– Add more instruction and training in communication, negotiation, and scoping skills since many students are engaging with community partners for the first time. |
| Through mentorship and EDI training, students learn to overcome challenges of working in a diverse team. | – Include explicit EDI training throughout the course of the project and emphasize the benefits of working with diverse teammates.<br>– Be prepared to provide mentorship to students as conflict is expected from time to time. |

# Bibliography

[1]  Pekka Abrahamsson, Outi Salo, Jussi Ronkainen, and Juhani Warsta. Agile software development methods: Review and analysis. Preprint at *arXiv:1709.08439*, 2017.

[2]  Sarah Andreas. Effects of the decline in social capital on college graduates soft skills. *Industry and Higher Education*, 32(1):47–56, 2018.

[3] William Aspray and Andrew Bernat. Recruitment and retention of underrepresented minority graduate students in computer science. In *Report on a Workshop by the Coalition to Diversity Computing*, 2000.

[4] Carlo Barone. Some things never change: Gender segregation in higher education across eight nations and three decades. *Sociology of Education*, 84(2):157–176, 2011.

[5] Virginia Braun and Victoria Clarke. Using thematic analysis in psychology. *Qualitative Research in Psychology*, 3(2):77–101, January 2006. Publisher: Routledge eprint: `www.tandfonline.com/doi/pdf/10.1191/1478088706qp063oa`.

[6] Linda B. Glaser. Research offers new hope for gender equity in STEM fields. 2017.

[7] Sujin K. Horwitz. The Compositional Impact of Team Diversity on Performance: Theoretical Considerations. *Human Resource Development Review*, 4(2):219–245, June 2005.

[8] David A. Kolb. *Experiential Learning: Experience as the Source of Learning and Development*. FT Press, 2014.

[9] Scott A. Lee. Increasing student learning: A comparison of students' perceptions of learning in the classroom environment and their industry-based experiential learning assignments. *Journal of Teaching in Travel & Tourism*, 7(4):37–54, 2008.

[10] Percival Lucena, Alan Braz, Adilson Chicoria, and Leonardo Tizzei. IBM design thinking software development framework. In *Brazilian Workshop on Agile Methods*, pages 98–109. Springer, 2016.

[11] Valerie Martinelli. What Happens When Your Soft Skills Kill Your Career? April 2019.

[12] Michael Miles, David Melton, Michael Ridges, and Charles Harrell. The benefits of experiential learning in manufacturing education. *Journal of Engineering Technology*, 22(1):24, 2005.

[13]   John Noll, Sarah Beecham, Abdur Razzak, Bob Richardson, Ann Barcomb, and Ita Richardson. Motivation and autonomy in global software development. In *International Workshop on Global Sourcing of Information Technology and Business Processes*, pages 19–38. Springer, 2017.

[14]   Polly S. Rizova. Are you networked for successful innovation? *MIT Sloan Management Review*, 47(3):49–55, 2006.

[15]   Kaela S Singleton, De-Shaine RK Murray, Angeline J. Dukes, and Lietsel NS Richardson. A year in review: Are diversity, equity, and inclusion initiatives fixing systemic barriers? *Neuron*, 109(21):3365–3367, 2021.

[16]   Bruce W. Tuckman. Developmental sequence in small groups. *Psychological Bulletin*, 63(6):384, 1965.